\documentclass{article}
\usepackage[cp1251]{inputenc}
\usepackage[russian]{babel}
\usepackage{graphicx}
\begin{document}
{Astrophysics, vol. 66, in press}
\rm
\let\sc=\sf
\begin{center}
\LARGE {\bf Photometric activity of CQ Tau on the time interval of 125 years}
\\
\vspace{10mm} \large {V.P.\,Grinin$^{1,2}$, L.V.\,Tambovtseva$^1$, О.Yu.\,Barsunova$^{1}$,
D.N.\,Shakhovskoy$^{3}$}
\end{center}
\vspace{0.6cm}
%\large
\normalsize
1 - Pulkovo Astronomical Observatory, Russian Academy of Sciences, Pulkovskoe sh. 65, St. Petersburg,
196140, Russia\\
2 - V.V. Sobolev Astronomical Institute, St. Petersburg State University, Universitetskii pr. 28, St. Petersburg,
198504, Russia\\
3 – Crimean Astrophysical Observatory, Russian Academy of Sciences, Crimea, Nauchny, Russia\\

\vspace{10mm}
%\Large
%\normalsize
\large {\bf Abstract}.\, The star CQ Tau belongs to the family of UX Ori type stars. It has very complex
photometric behavior and complex structure of the circumstellar environment. In our paper we constructed the
historical 125 years light curve of this star basing on the published photometric observations. It follows that
besides a random component characteristic of UX Ori type stars, the large amplitude periodic component with the 10
year period is also present. Its existence was suspected earlier in [11]. New observations confirm its reality. It
points to an existence of the second component close to the star. The density waves and matter flows caused by the
companion motion lead to periodic changes in the circumstellar extinction and brightness of the star. This result
is discussed in context of the recent observations of CQ Tau with high angular
resolution. \\

\clearpage
\section{Introduction}
The star CQ Tau (Sp = F5 IVe, Mora et al. [1]) is one of the most active UX Ori type stars (UXORs). Its brightness
varies with an amplitude up to $\Delta V\approx 3^m$. The star demonstrates all signatures typical for UXORs.
Among them the so-called “blueing” effect is present consisting in the shift of the star color to the blue part of
the spectrum in the deep brightness minima. This effect was firstly observed namely in CQ Tau (Gotz \& Wenzel
[2]), and its first interpretation was based on the assumption that this star was a binary and had a weak blue
companion (Wenzel [3]). When the main component is screened by the circumstellar (CS) dust cloud the radiation of
the blue companion begins to dominate. We will return to the idea of the CQ Tau binarity later, and now let us
note that discovery of the same “blueing” effect in other UXORs closed the idea of binarity as a possible reason
of this effect. The current explanation of this effect assumes that the blue radiation source of UXORs is the
scattered radiation of the protoplanetary disks whose contribution increases during deep minima [4]. Observations
of the high linear polarization in the deep minima confirmed this model (see [5] and cited papers therein). On the
base of these observations it has been suggested that the CS disks of UXORs are inclined at a small angle to the
line of sight, and this is the main reason of their specific variability.

This conclusion was generally supported with interferometric observations (see Kreplin et al. [6]) and cited
papers therein). However, in the case of CQ Tau the situation was more complicated. Interferometry in the near
infrared (IR) spectrum region revealed (Eisner et al. [7]), that the inclination angle of the inner region of the
CS disk to the line of sight was equal to 48$^\circ$ that contradicted to the status of UXORs. Interferometric
observations in the submillimeter range of the spectrum turned out to be even more surprising. They showed that
the outer part of the disk is observed almost pole-on (Chapillon et al. [8], and this result was recently
confirmed by observations with the ALMA interferometer (Ubeira-Gabellini et al. [9]).

No less complex is also the long-term brightness variability of the star (Minikulov et al. [10], Shakhovskoy et
al. [11], Grinin et al. [12]). Using the data from these papers, we can trace the photometric activity of CQ Tau
during about of 100 years. The first observations were fulfilled with the photographic method. They showed that
the star was bright for a long time (about of 40 years), and its brightness fluctuated within 0.4-0.7m [12].
Around the middle of the last century the photometric activity of CQ Tau has changed dramatically: the star began
to demonstrate deep brightness minima with an amplitude up to 2-3$^m$. The long lasting cycles began to be
observed along with stochastic variability.

Periodogram analysis of the photometric series revealed two large periods: one with duration of about 20-21 years
and another period was about half of the first one: $\approx$ 10 years [11]. After their removing a short 3 year
period has been found.

Taking into account the large duration of the activity cycles, new observations are needed for their confirmation.
The last observation of the CQ Tau, used in [11,12], was fulfilled in 2003. Twenty years passed since that time,
and now a possibility appeared to prolong the brightness curve of the star. We consider below what does it give
for study of the photometric activity of CQ Tau.

\section{Historical light curve of CQ Tau}
Figure 1 shows the light curve of the star in the B band plotted according to the data of [10-12]. It is completed
with new observations from the ASAS-SN (All-Sky Automated Survey for SuperNovae, Kochanek et al., [13]) and AAVSO
(the American Association of Variable Star Observers: https://www.aavso.org/) databases. When we plotted this
light curve we used published observations fulfilled both by cited authors and by authors to whom they referenced
in their papers. As it is known, observations presented in ASAS-SN as well as the most of observations in AAVSO
were made in the V band. In order to transform them to the B band we used observations by Berdugin et al. [14]
made in UBVRI bands. With their help we obtain a function connected magnitudes of CQ Tau in the B and V bands B =
- 0.18V2 + 4.73V - 18.54. As it is shown in Appendix, this ratio ensured accuracy 5\% for transition from V to B,
that is quite enough for our purposes taking into account that changes in the brightness amplitude are 3
magnitudes.

\begin{figure}
\centering
\includegraphics[angle = 0, width=15cm]{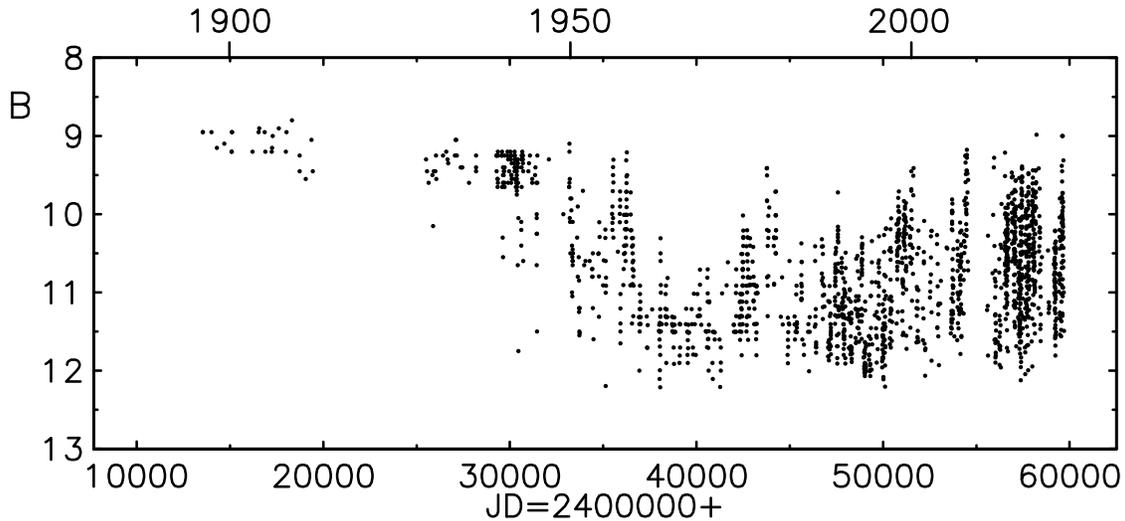}
\caption{Historical light curve of CQ Tau in the B-band according to the published data. The first observation has
been done in Moscow in 1895.}
\end{figure}
It is directly seen from the CQ Tau light curve that the large cycle of the photometric activity with duration of
20-21 year is not traced in the new observations. On the contrary, the 10 year cycle is clearly seen. It is also
confirmed by the periodogram analysis of the photometrically most active part of the star light curve (MJD >
35000) shown in Fig. 2. One can see a presence of two periods: 10 years and 321.1 days. The latter is the
year-linked to the 10-year period, it reflects the presence of the annual breaks in the observations.
\begin{figure}
  \centering
   \includegraphics[angle = 0, width=10cm]{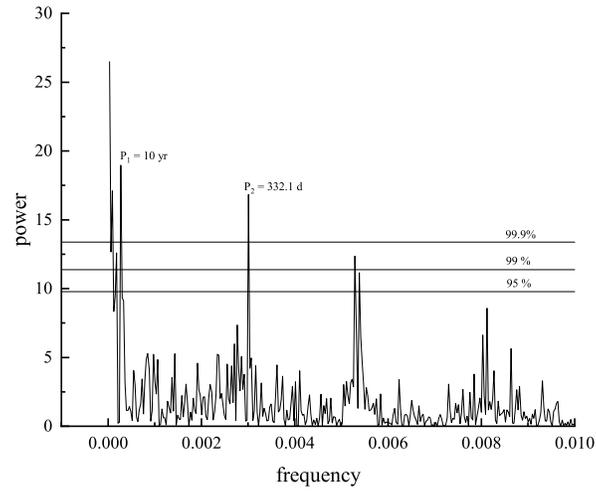}
  \caption{The Lomb-Scargle periodogram of the most active part (JD > 35000) of the CQ Tau
photometric series.}
\end{figure}
After subtracting the 10 year period we did not find the traces of the 3 year period obtained by Shakhovskoy et
al. [11] in the remnant periodogram. Thus, the new periodogram analysis of the CQ Tau photometric activity
confirmed a reality of only 10 year period. The convolution of the investigated part of the photometric series
with the 10 year period is shown in Fig. 3.

From Fig. 3 it follows that the 10 year period is mainly revealed in the periodic modulation of the brightness at
the bright state of the star and amplitudes of the minima. As it is seen from the light curve (Fig. 1), this
modulation is observed on the systematic increase in the attenuation amplitude. Also the gradual brightening trend
is distinctly seen.
\begin{figure}
  \centering
    \includegraphics[angle = 0, width=10cm]{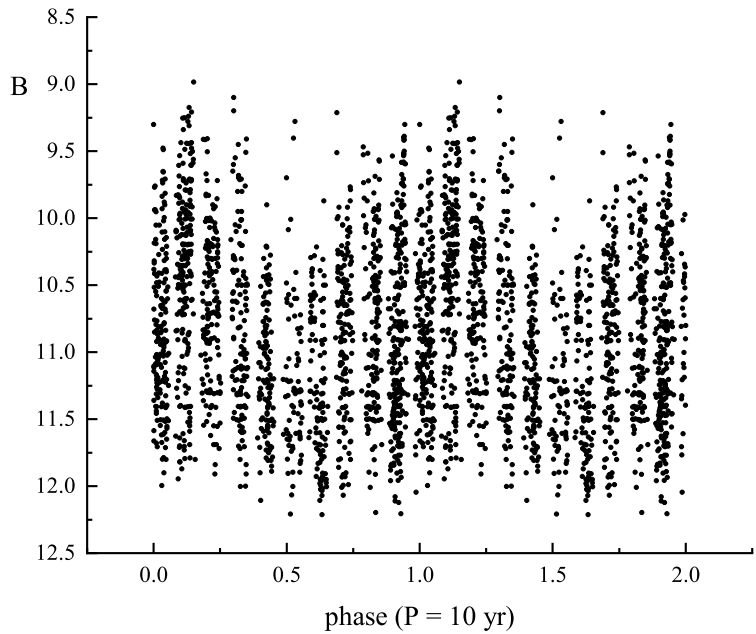}
  \caption{The convolution of the CQ Tau observational photometric series with the 10 year period.}
\end{figure}
\section{Discussion and Conclusion}
Presence of the 10-years period of the photometric activity in CQ Tau indicates to an existence of the companion
in the vicinity of the star. Results of the interferometric observations of CQ Tau in the millimeter range testify
the same (Tripathi et al. [15]; Wolfer et al. [16]; Ubeira-Gabellini et al. [9]). They showed that there is a vast
cavity weakly filled with the matter in the central part of the protoplanetary disk of the star. Such cavities are
formed in the young binary systems due to the tidal disturbances caused by the orbital motion of components
(Artymowicz and Lubow [17]). Periodic perturbations lead to the formation of spiral density waves, and they are
actually observed in images of the CQ Tau disk (Ubeira-Gabellini et al. [9], Uyama et al. [18], Hammond et al.
[19], Safonov et al. [20]. However attempts to find a companion have not been successful so far [19].

According to Ubeira-Gabellini et al. [9], the cavity in the CQ Tau protoplanetary disk stretches from 15 to 25 AU.
Numerical simulation performed by these authors showed that such a cavity can be formed by the planet with the
mass of 6-9 M$_{Jup}$ moving on the circular orbit with the radius of 20 AU. The orbital period of the planet with
the mass of 1.67 M$_\odot$ will be equal to 69 years that obviously contradicts to the 10 year photometric period.

Is there a possibility to avoid this contradiction? We suppose that such a possibility exists if to drop an
assumption about the circular orbit of the secondary and to increase its mass. According to models by Artymowicz
and Lubow [17], in this case one can obtain the cavity of the same size with at a lower value of the large
semi-axis of the orbit but respectively with a lower period. One should keep also in mind that according to the
interferometric data in the near IR spectrum region (Eisner et al. [7]) the inner disk of CQ Tau is inclined
relatively to the outer disk. This means that the orbit of the low mass companion can be also inclined relatively
to the plane of the main disk. We intend to consider these questions in more details in the forthcoming paper. The
main question also remains open: what happened in the environment of the star, which led to radical change in its
photometric activity in the middle of the last century? \\

We thank E.N. Kopatskaya for useful comments. This work was supported by the Ministry of Science and Education of
the Russian Federation (project no. 075-15-2020-780).

\clearpage %\vspace{1cm}
\begin{center}
\LARGE {\bf References}
\end{center}
1.\, A.\,Mora, B.\,Merin, E.\,Solano, et al. Astron. Astrophys. \textbf{378}, 116 (2001).\\
2.\, W. von Gotz, and W. Wenzel, Mitt. Verand. Sterne, \textbf{5}, 2 (1968).\\
Wenzel, W., in L.Detre (ed.), “Non-Periodic Phenomena in Variable Stars”, IAU Colloq. (Budapest: Acad. Press), p.
61 (1969).\\
4.\, V.P. Grinin, Sov. Astron. Lett. \textbf{14}, 27 (1988).\\
5.\, V.P. Grinin, N.N Kiselev, N.K. Minikulov, G.P. Chernova, N.V. Voshchinnikov, Ap. Sp. Sci, \textbf{186},
283 (1991).\\
6.\, A. Kreplin, D.I. Madlener, L. Chen, et al. Astron. Astrophys. \textbf{590}, A96 (2016).\\
7.\, J.A. Eisner, B.F. Lane, L.A. Hillenbrand, R.L. Akeson, and A.I. Sargent, Astrophys. J. \textbf{613},
1049 (2004).\\
8.\, E. Chapillon, S. Guilloteau, A. Dutrey, et al.), Astron. Astrophys. \textbf{488}, 565 (2008).\\
9.\, M.G. Ubeira Gabellini, A. Miotello, S. Facchini  et al., MNRAS, \textbf{486}, 4638 (2019).\\
10.\, N. Kh. Minikulov, V. Yu. Rakhimov, N. A. Volchkova, and A. I. Pikhun, Astrophysics, \textbf{36}, 31 (1993).\\
11.\, D. N. Shakhovskoi, V. P. Grinin, and A. N. Rostopchina, Astrophysics, \textbf{48}, 135 (2005).\\
12.\, V.P. Grinin, O.Yu. Barsunova, S.Yu. Shugarov, P. Kroll, and S.G. Sergeev, Astrophysics, \textbf{51}, 1 (2008).\\
13.\, C.S. Kochanek, B.J. Shappee, K.Z. Stanek, et al. Publ. Astron. Soc. Pac., 129 (980) (2017), Article 104502\\
14.\, A.A. Berdyugin, S. V. Berdyugina, V. P. Grinin, and N. Kh. Minikulov, Soviet Astr. \textbf{34}, 408 (1990).\\
15.\, A. Tripathi, S.M. Andrews, T. Birnstiel, D.J. Wilner. Astrophys. J., \textbf{845}, 44 (2017).\\
16.\, L. Wolfer, S. Facchini, N.T. Kurtovic, et al., Astron. Astrophys. \textbf{648}, A19 (2021).\\
17.\, P. Artymowicz, S.H. Lubow, Astrophys. J., \textbf{421}, 651 (1994).\\
18.\, T. Uyama, T. Muto, D. Mawet, et al. Astron. J., 159, 118, \textbf{159}, 118 (2020).\\
19.\, I. Hammond, V. Christiaens, D.J. Price et al. MNRAS, 515, 6109 (2022).\\
20.\, B.S. Safonov, I.A. Strakhov, M.V. Goliguzova, O.V. Voziakova. Astron. J., 163, 31 (2022).\\

\section{Appendix}
As mentioned above, the main part of photometric observations of CQ Tau from ASAS and AAVSO archives were
fulfilled only in the V band. We determined relevant factors to transform the CQ Tau light curve from the V band
to the B band, and used them for plotting the historical light curve. For this purpose we used the photometric
observations of CQ Tau from the paper by Berdyugin et al. [13]. They are shown in Fig. 4. The dashed line shown in
the same figure is defined by the second degree polynomial obtained when fitting to observations with the least
square method.
\begin{figure}
  \centering
  \hspace{0cm}
  \includegraphics[angle = 0, width=10cm]{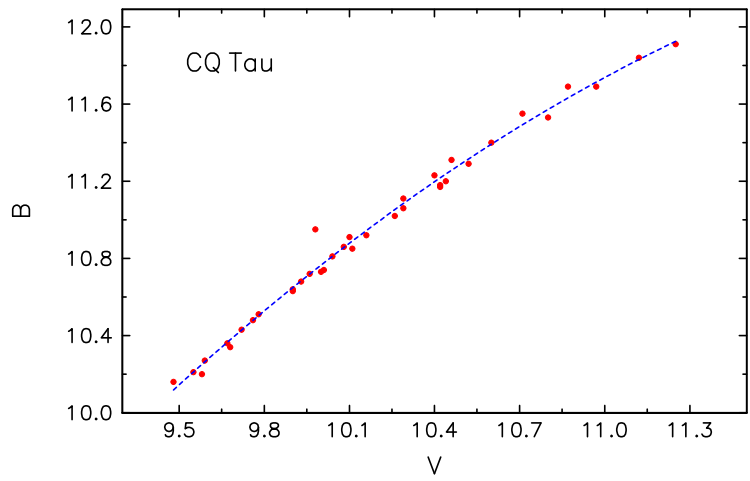}
  %\vspace{-4cm}
  \caption{CQ Tau stellar magnitudes in the B and V bands according to data by Berdyugin et al. [13].}
\end{figure}
The dashed line in Fig. 4 sets a functional link between B and V values and described by the ratio: B = - 0.18
V$^2$ + 4.73V - 18.54. One can see that an accuracy of the V to B transition with the use of this ratio no worse
than 0.05$^m$ for most observations presented in Fig. 4.
\end{document}